\documentclass[10pt,a4paper,onecolumn]{article}

\usepackage[journal=jpclcd]{achemso}
\setkeys{acs}{articletitle = true}
\usepackage[T1]{fontenc}       
\usepackage{graphicx}
\usepackage{dblfloatfix}
\usepackage{placeins}
\usepackage{authblk}
\usepackage{a4wide}
\usepackage{setspace}
\usepackage[margin=1in]{geometry}
\usepackage{physics}
\usepackage{color}
\usepackage{bm}
\usepackage{booktabs, tabularx} 
\usepackage{mathrsfs} 
\newcolumntype{C}{>{\centering\arraybackslash}X}

\author[1]{Amitav Sahu}
\author[1]{Vivek Tiwari \thanks{vivektiwari@iisc.ac.in}}
\affil[1]{Solid State and Structural Chemistry Unit, Indian Institute of Science, Bangalore, Karnataka 560012, India}

\newcommand*{\vt}[1]{\textcolor{black}{ #1}}
\newcommand*{\AS}[1]{\textcolor{black}{ #1}}
\newcommand*{\add}[1]{\textcolor{black}{ #1}}
\begin{document}

\title {Coherence Transfer and Destructive Interference in Two-Dimensional Coherence Maps}
\maketitle

\begin{abstract}
Coherence maps (CMs) in multidimensional spectroscopy report total interference of all quantum coherent pathways. Detailed understanding of how this interference manifests spectroscopically is vital for deciphering mechanistic origins of impulsively generated wavepackets, but currently lacking.  Here we explain the origin of recently reported diagonal node-like features in CMs of \textit{bacteriochlorophyll} monomers and photosynthetic reaction centers (RCs), where the apparent resemblance in the two disparate systems was reportedly perplexing. We show that both spectroscopic signatures have distinct physical origins. Node-like lineshapes in monomers arise from unique phase twists caused by destructive interference between ground and excited state vibrational coherences. In contrast, nodal lines in RCs are explained by coherence transfer of vibrational wavepackets which do not participate in the ultrafast energy transfer and their destructive interference with ground state pathways. Our results resolve recent spectroscopic observations and illustrate new mechanistic insights gained from understanding interference effects in multidimensional spectroscopy.
\end{abstract}

Two-dimensional electronic spectroscopy\cite{JonasARPC2003} (2DES) resolves ultrafast dynamics ensuing femtosecond excitation as 2D contour map snapshots along initially correlated excitation and detection frequency axes, evolving along the pump-probe waiting time $T$. Analysis of coherent signal contributions, often accompanying impulsive excitations, along the corresponding frequency axis $\omega_T$ leads to coherence maps (CMs). At a given $\omega_T$, CMs report the total interference of all quantum coherent Feynman pathways which may arise from distinct origins, for instance, from purely vibrational versus mixed vibrational-electronic wavepackets, or from wavepackets on ground versus excited electronic states. \\

CM analysis of peak positions\cite{Thyrhaug2018} can disentangle overlapping signal contributions with varying degree of success, and despite spectral decongestion along three dimensions, ambiguous spectroscopic signatures can still arise. For example, recent 2DES experiments on photosynthetic reaction centers (RCs) have reported\cite{Palecek2017,PolichtThesis} diagonal nodes in CMs for all reported intramolecular vibrational frequencies. However similar diagonal node-like interference effects were later reported\cite{Policht2018} in CMs of bacteriochlorophyll (\textit{BChl a}) monomers. These similarities in spectroscopic signatures from multichromophoric RCs and \textit{BChl a} monomers in solution were reportedly perplexing. Here we show that these apparently similar spectroscopic signatures arise from distinct physical mechanisms, and illustrate how CM lineshapes serve as subtle reporters of the underlying physics of vibrational coherence transfer\cite{Jean1995,Cina2004} and destructive interference between signal pathways contributing to 2D spectra.

\begin{figure*}[h!]
	\centering
	\includegraphics[width=3.25 in]{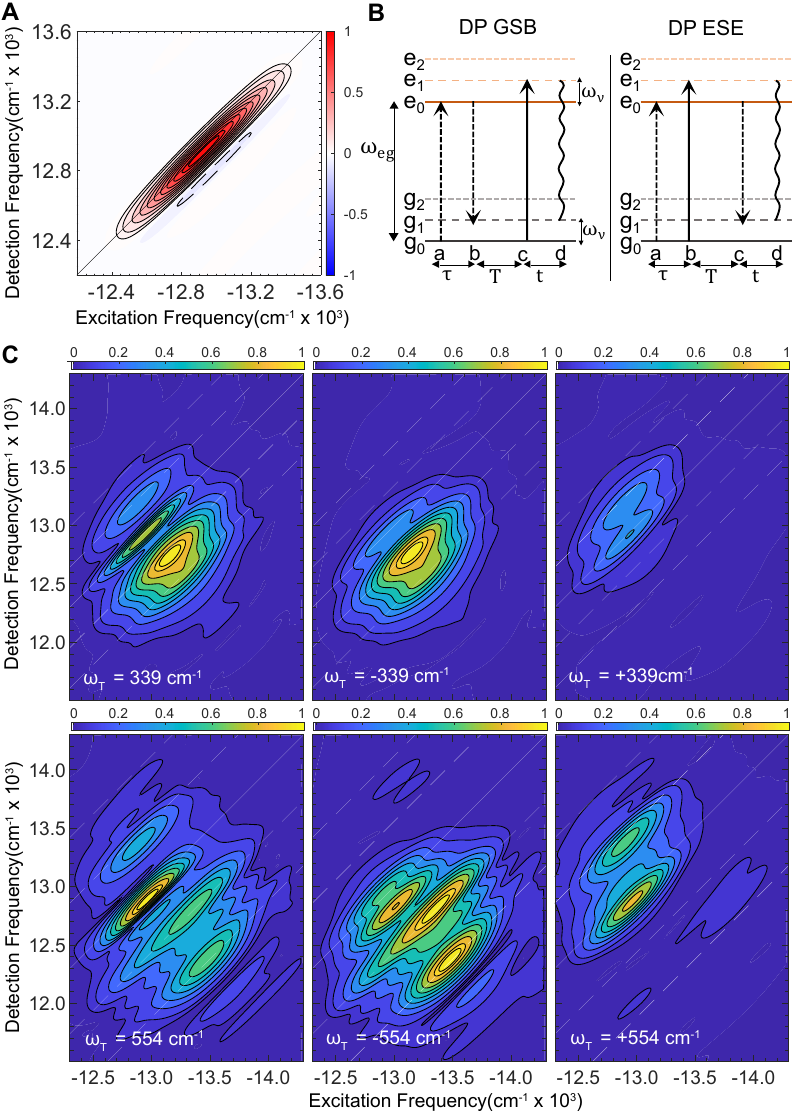}
	\caption{\footnotesize (A) Real absorptive $T=50$ fs 2D spectrum of a two electronic level system with \AS{five} underdamped FC vibrations modeled as Brownian oscillators to simulate coherence maps (CMs) for dominant vibrational coherences recently reported\cite{Policht2018} in \textit{Bacteriochlorophyll a} monomer. The spectrum is calculated at 77 K. The model parameters are described in \AS{Table S2}. Contours are drawn at the  5$\%$, and 10-90$\%$ in 10$\%$ intervals for positive or negative contours. (B) Wavemixing diagrams corresponding to 2DES signal contributions at the diagonal peak arising from ground (GSB) and excited state (ESE) vibrational coherences. (C) Real rephasing (left) and -$\omega_T$  (center) and +$\omega_T$ (right) complex rephasing CMs for vibrational frequencies 339 cm$^{-1}$ and 554 cm$^{-1}$. \AS{CMs at other vibrational frequencies are shown in Fig.~S4}. The narrowing of diagonal CM lineshape in the real rephasing CMs due to the destructive interference between overlapping GSB and ESE coherent pathways of panel B is evident for all vibrational coherences in the model.}
	
	\label{fig:fig1}
\end{figure*}

To address the above questions, we start by deriving analytic expressions for CM lineshapes to show that uniquely different phase-twists, as opposed to those arising from imbalanced rephasing and non-rephasing signals, can arise due to interference between Feynman pathways for ground and excited vibrational wavepackets. We will show that reported\cite{Policht2018} diagonal node-like features in 2DES CMs of \textit{BChl a} monomers are manifestations of this interference. We consider the simplest model of a three electronic level system with a Franck-Condon (FC) active intramolecular vibration identically coupled to all three electronic states. \vt{Note that unlike the transition strengths derived from 2-electrons in a 2D box model for D$_{4h}$ symmetric monomers, only one-electron transition strengths are expected\cite{Weiss1972} in \textit{BChl a} due large electronic splitting between the Q$_{x}$ and Q${_y}$ bands}. The ground state bleach (GSB), excited state emission (ESE) and absorption (ESA) Feynman pathways for vibrational quantum coherences that contribute to the 2DES diagonal peak (DP) can be\vt{written as a product of an orientational factor arising from four transition dipole factors interacting with the pump and probe electric fields, and a Green's function time-propagator $\mathscr{G}(t)$ for each time interval between light-matter interactions.} The GSB, ESE and ESA rephasing Feynman pathways for vibrational coherences on the 2D diagonal are given by \AS{Eqns.~S1--S3} and represented as wavemixing diagrams in \vt{Fig.~\ref{fig:fig1}B and Fig.~S1}.

Only dominant $0-1$ vibrational coherences and Bloch dephasing have been considered in \AS{Eqns.~S1--S3} for the purpose of deriving analytic expressions. In the Bloch limit, $\mathscr{G}_{mn}(t) = \theta(t)exp[-\gamma_{mn}(t)]exp[-i\omega_{mn}t]$ , where $\theta(t)$ is the Heaviside step function, $\omega_{mn} = (E_m - E_n)/\hbar$, and $\gamma_{mn}$ is the dephasing rate.  It is reasonable to expect ground and excited state vibrational coherences to dephase with different rates along $T$, denoted by $\gamma_{g}$ and $\gamma_{e}$, respectively. Optical Bloch dephasing rates have been assumed to be equal for simplicity, and denoted by $\gamma$. \vt{Note that the 2D lineshape is determined by the product of Green functions, while the transition dipoles impart an overall strength and sign. All such individual lineshapes interfere to result in total 2D signal strength and lineshape.} Crucially, sign of the coherence frequency along $T$ is opposite for GSB versus ESE and ESA pathways (Fig.~S1), and dictates the diagonal node-like interference feature as explained below. \\

Inverse Fourier transformation of \AS{Eqns.~S1-S3} along the first and third time intervals yields 2D lineshapes along excitation and detection frequencies, $-\omega_{\tau}$ and $\omega_t$, for each non-zero value of waiting time $T$. For the case of GSB pathways, the resulting frequency domain complex 2D signal $\tilde{S}_3$ is given by -- 
	\begin{eqnarray}
	\tilde{S}_{3}(-\omega_\tau,\omega_t;T) = \mathscr{F}^{-1}[\mathscr{G}_{g_0e_0}(\tau)] \mathscr{G}_{g_0g_1}(T) \mathscr{F}^{-1}[\mathscr{G}_{e_1g_1}(t)] \nonumber\\
	= [\alpha(-\omega_{\tau},\omega_t) - i\beta(-\omega_{\tau},\omega_t)]e^{-i\omega_{v}T}e^{-\gamma_{g}T},
	\label{eq2}
\end{eqnarray}
where vibrational coherence frequency along $T$ has been substituted by the vibrational frequency $\omega_v$. $a_{mn}(\omega)$ and $d_{mn}(\omega)$ are absorptive and dispersive 2D Lorentzian lineshapes, respectively, with
\begin{eqnarray}\label{eq3}
\alpha (-\omega_{\tau},\omega_t) = [a_{g_0e_0}(-\omega_\tau) a_{e_1g_1}(\omega_t) - d_{g_0e_0}(-\omega_\tau) d_{e_1g_1}(\omega_t)]  \nonumber \\
\beta(-\omega_{\tau},\omega_t) = [a_{g_0e_0}(-\omega_\tau) d_{e_1g_1}(\omega_t) +a_{e_1g_1}(\omega_t) d_{g_0e_0}(-\omega_{\tau})]
\end{eqnarray}
In the same fashion, including the ESE and ESA contributions, the total complex 2D signal becomes --
\begin{eqnarray}
	\tilde{S}_{tot}(-\omega_\tau,\omega_t;T) = 	 [\alpha(-\omega_{\tau},\omega_t) - i\beta(-\omega_{\tau},\omega_t)] (e^{i\omega_{v}T}e^{-\gamma_{e}T}(1-\kappa) + e^{-i\omega_{v}T}e^{-\gamma_{g}T}).
	\label{eq4}
\end{eqnarray}
where an ESA strength factor $\kappa$ has been included to account for any differences in transition strengths of doubly-excited states. \AS{See Section S1 for details of the derivation}. To derive CM lineshapes from a real rephasing 2D spectrum $S^R_{tot}(-\omega_\tau,\omega_t;T)$, we will consider special cases of Eqn.~\ref{eq4}. Experimental absorptive 2D spectra of \textit{BChl a} monomers have reported\cite{Policht2018} very weak off-diagonal contributions from excited state absorption (ESA) with negligible contribution on the 2D diagonal. Accordingly we simplify Eqn.~\ref{eq4} to first consider only excited state emission (ESE) and ground state bleach (GSB) pathways with $\kappa$ set to zero. Assuming dephasing rates of excited and ground state vibrational coherences such that $\gamma_e \approx \gamma_g = \gamma_{g,e}$, and Fourier transforming along $T$ yields the corresponding CM lineshape contributing on the diagonal with frequency $\omega_T = \omega_{v}$, $CM_{tot}(-\omega_\tau,\omega_t;\omega_T) = \text{abs}[\alpha(-\omega_{\tau},\omega_t) a(\omega_T = \omega_v)]$, where the absolute value is consistent with how 2DES CMs are typically reported.
 Inspection of the CM lineshape $CM_{tot}(-\omega_\tau,\omega_t;\omega_T)$ suggests reduction in diagonal amplitude due to cancellations between $a(-\omega_\tau) a(\omega_t)$ and $d(-\omega_\tau) d(\omega_t)$ terms in $\alpha(-\omega_{\tau},\omega_t)$ (Eqn.~\ref{eq4}). Although the latter term vanishes at the peak center, it removes amplitude from off-center locations above and below the diagonal. This destructive interference between ESE and GSB coherence pathways on the DP is distinct from the phase-twist quantum beats arising\cite{Cho2009} from imbalance between rephasing and non-rephasing pathways. The GSB/ESE destructive interference discussed here only arises at the DP and not on other 2D CM locations (\AS{see Fig.~S2}) because opposite phases of vibrational quantum coherences only overlap on the diagonal (Eqn.~\ref{eq3}). 2DES simulations with Bloch lineshapes confirm the narrowed diagonal lineshapes in CMs arising from destructively interference between GSB/ESE vibrational coherence pathways. \AS{These are shown in Section S1.}

\FloatBarrier

We extend the above analytic reasoning to simulate the recently reported 2D CMs of \textit{BChl a} monomers at 77K. The model parameters are described in \AS{Section S2}. Briefly, the reported FC active intramolecular vibrations are modeled as underdamped Brownian oscillators with stabilization energies, damping and frequencies similar to those reported for \textit{BChl a} monomers\cite{Policht2018,Frieberg2011,Wendling2000}.  Energetic disorder of \AS{230} cm$^{-1}$ in the optical energy gap is also included in the model to approximately match the diagonal linewidth reported\cite{Policht2018} for early $T$ 2DES spectra of \textit{BChl a} monomers. \AS{All the simulation parameters are summarized in Table S2.} Figure \ref{fig:fig1}A shows the $T=50$ fs absorptive 2DES spectrum. \AS{A Frobenius spectrum of vibrational coherences corresponding to all intramolecular vibrations in the model is shown in Fig.~S3.} Fig.~\ref{fig:fig1}B shows the GSB and ESE wavemixing diagrams which interfere on the 2D diagonal. The real rephasing CMs for two of the intramolecular vibrations are shown in the left column of Fig.~\ref{fig:fig1}C. \AS{Rest of the CMs are shown in Fig.~S4}. As expected, for all vibrations, the  $d_{g_0e_0}(-\omega_\tau) d_{e_1g_1}(\omega_t)$ term in $CM_{tot}(-\omega_\tau,\omega_t;\omega_T)$, that results from the interference of GSB and ESE vibrational coherence pathways, leads to narrow node-like features on the 2D diagonal. Compared to simulations with Bloch lineshapes (\AS{Fig.~S2}), these features are further accentuated by energetic disorder in the optical energy gap in the reported\cite{Policht2018} 2D spectra at 80K. It can be easily verified that when one starts from the complex rephasing 2D spectrum $\tilde{S}_{tot}(-\omega_\tau,\omega_t;T)$ (Eqn.~\ref{eq4}), the absolute value CM lineshape for the case of $\omega_T = \pm\omega_{v}$ is now given by $\mathscr{R}(-\omega_\tau,\omega_t) = \sqrt{\alpha^2(-\omega_\tau,\omega_t;T) + \beta^2(-\omega_\tau,\omega_t;T)}$, which does not predict a diagonal narrowing but instead an approximately absorptive lineshape. It is therefore no surprise that when the CMs are resolved according to the quantum beat phase $\pm\omega_T$ (Fig.~\ref{fig:fig1}C middle and right panels), DP narrowing due to GSB/ESE destructive interference does not occur. Note that the above reasoning suggests that the diagonal node-like feature is not specific to \textit{BChl a}. Recent real rephasing CMs for Oxazine 720 monomers\cite{Sahu2023} are consistent with this expectation.\\

In deriving $CM_{tot}(-\omega_\tau,\omega_t;\omega_T)$ lineshape it is assumed that the dephasing timescale of excited and ground state vibrational coherences is comparable, that is, $\gamma_g \sim \gamma_e$. However, ultrafast electronic relaxation channels can cause large anharmonicities on the excited state potentials such that some excited state vibrational wavepackets may not survive electronic relaxation\cite{JonasARPC2018,Scholes2019Review}. For example, vibrational quantum beats from `promoter' modes\cite{Patra2021} that \textit{tune} the relative energy gaps and electronic couplings between excited state potentials do not survive relaxation though a conical intersection and dephase\cite{Jonas2008} within $\sim$100 fs. In contrast, the ground state beats survive for picoseconds. The above assumption will also be invalidated in systems exhibiting ultrafast singlet exciton fission, such as pentacene thin films, due to rapid internal conversion\cite{Friend2011} of excited state population into correlated triplet states. In all such cases, Eqn.~\ref{eq4} predicts absence of diagonal node-like GSB/ESE interference feature in the CMs. 

Interestingly, the CM lineshape then also becomes a useful spectroscopic reporter of `promoter' vibrational modes and excited state dynamics. For example, Policht et al. reported\cite{Policht2018} no change in the diagonal node-like feature for \textit{BChl a} for penta- or hexa-coordinating solvents, suggesting no substantial effect on the dephasing rates of ground and excited vibrational wavepackets due to solvent coordination. Recent broadband pump-probe measurements from Scholes and coworkers report that solvent tuning of electron transfer rates can dephase\cite{YonedaScholes2021} vibrational modes parallel to the reaction coordinate up to 5x faster, whereas `spectator' modes remain unaffected. Our analysis of GSB/ESE interference predicts that in a corresponding 2DES experiment, as opposed to `spectator' modes, the diagonal CM lineshapes for `promoter' modes will \textit{not} show the node-like diagonal lineshape. 


Having explained the node-like features reported for \textit{BChl a} monomers, we can now investigate similar spectroscopic signatures that were reported\cite{Palecek2017,Policht2022} to accompany ultrafast energy transfer in \add{bacterial} RC proteins \add{(BRCs)}. Recently, Zigmantas and co-workers reported\cite{Palecek2017} diagonal nodes in all the vibrational CMs from a chemically oxidized BRC undergoing sub-200 fs $H \rightarrow B$ energy transfer. \add{Interestingly, similar nodes are also reported by Ogilvie and co-workers\cite{Policht2022,PolichtThesis} in the context of $B \rightarrow P$ energy transfer for mutant BRCs, even when charge separation is precluded\cite{Niedringhaus2018} ($D_{LL}$ mutant). Similarities in diagonal nodes for the two cases are curious. Furthermore, in all above cases, dominant ESA contributions on the upper diagonal 2D peak ($DP_U$) are also reported. Bukart\.{e} et al. have explained\cite{Zigmantas2021} these contributions using ESA related "re-excitation" Feynman pathways with dispersive lineshapes caused by electrochromic shifts induced when charge separation is complete at long waiting times pump-probe ($T > $2 ps). However, dispersive lineshapes are seen\cite{Niedringhaus2018} for as early as 250 fs, and even for the $D_{LL}$ mutant. For example, see Fig.~S2 and Fig.~S6 of ref. \cite{Niedringhaus2018}}. \\

\add{Pale\v{c}ek et al.\cite{Palecek2017} have qualitatively argued for an excited to ground state coherence shift explain the nodal feature. Recent experiments, supported by simulations, from Policht et al.\cite{Policht2022} explain coherent ESA contributions in the upper 2D cross-peak region by incorporating a distinctly different coherence transfer mechanism between excited state vibronic eigenstates. Overall, presence of nodal lines on the diagonal for all reported intramolecular vibrations, even for BRC mutants incapable of charge separation, and the similarity with node-like diagonal features in \textit{BChl a} monomers\cite{Policht2018} has remained perplexing and begs further explanation. While destructive GSB/ESE interference (Fig.~\ref{fig:fig1}) already explains the diagonal node-like CM lineshapes in monomers, below we explain the distinct physical origin for the reported diagonal nodal lines.} \\




We consider an excitonic dimer model with two intramolecular FC vibrations, as a minimum model for $P-B$ or $B-H$ exciton pairs studied earlier\cite{Palecek2017,Policht2022}. The rapid energy transfer process is complete within $\sim$200 fs and the reported CMs correspond to vibrational wavepackets which survive energy transfer. In order to incorporate electronic relaxation in wavemixing pathways, \vt{we adopt the recently reported\cite{Engel2022} approach of Engel and co-workers which exploits symmetries between 2D diagonal and cross peaks (as they grow or decay due to electronic relaxation) to extract population transfer kinetics.} In the context of coherence transfer accompanying ultrafast electronic relaxation, multilevel Redfield simulations\cite{Jean1995} of Jean and Fleming are quite instructive. Their results suggest that coherent vibrational motions orthogonal to the `reaction coordinate' can undergo coherence transfer to the acceptor through dominant secular terms in the Redfield tensor, $\mathcal{\textbf{R}}_{\alpha \beta, \gamma \delta}$ as long as the coherence frequency is maintained on the donor and acceptor excitons, that is, $\omega_{\alpha \beta} = \omega_{\gamma \delta}$, respectively. This is so because orthogonal spectator motions maintain the donor-acceptor energy gap, while energy gap tuning motions lead to `nesting'\cite{Scholes2019Review} of donor-acceptor electronic states and may not survive energy transfer. Witkowskii and Moffitt\cite{Moffit1960}, and later others\cite{Gouterman1961,Sinanoglu} have analyzed donor-acceptor energy transfer in terms of symmetric or correlated $\hat{q}_+$ and anti-symmetric or anti-correlated $\hat{q}_-$, relative motions of intramolecular vibrational coordinates $\hat{q}_{A,B}$ on the respective molecules. \vt{Using a surface-crossing description of electronically coherent donor-acceptor energy transfer, Cina and Fleming\cite{Cina2004} have elucidated the interplay of vibrational coherence transfer and in-phase or symmetric vibrational motions between the donor and acceptor.} Correlated vibrations play the role of spectator motions in an excitonic dimer. Since they maintain donor-acceptor energy gap, coherence transfer is expected to be dominant along correlated modes.  

\begin{figure*}[h!]
	\centering
	\includegraphics[width=3.25 in]{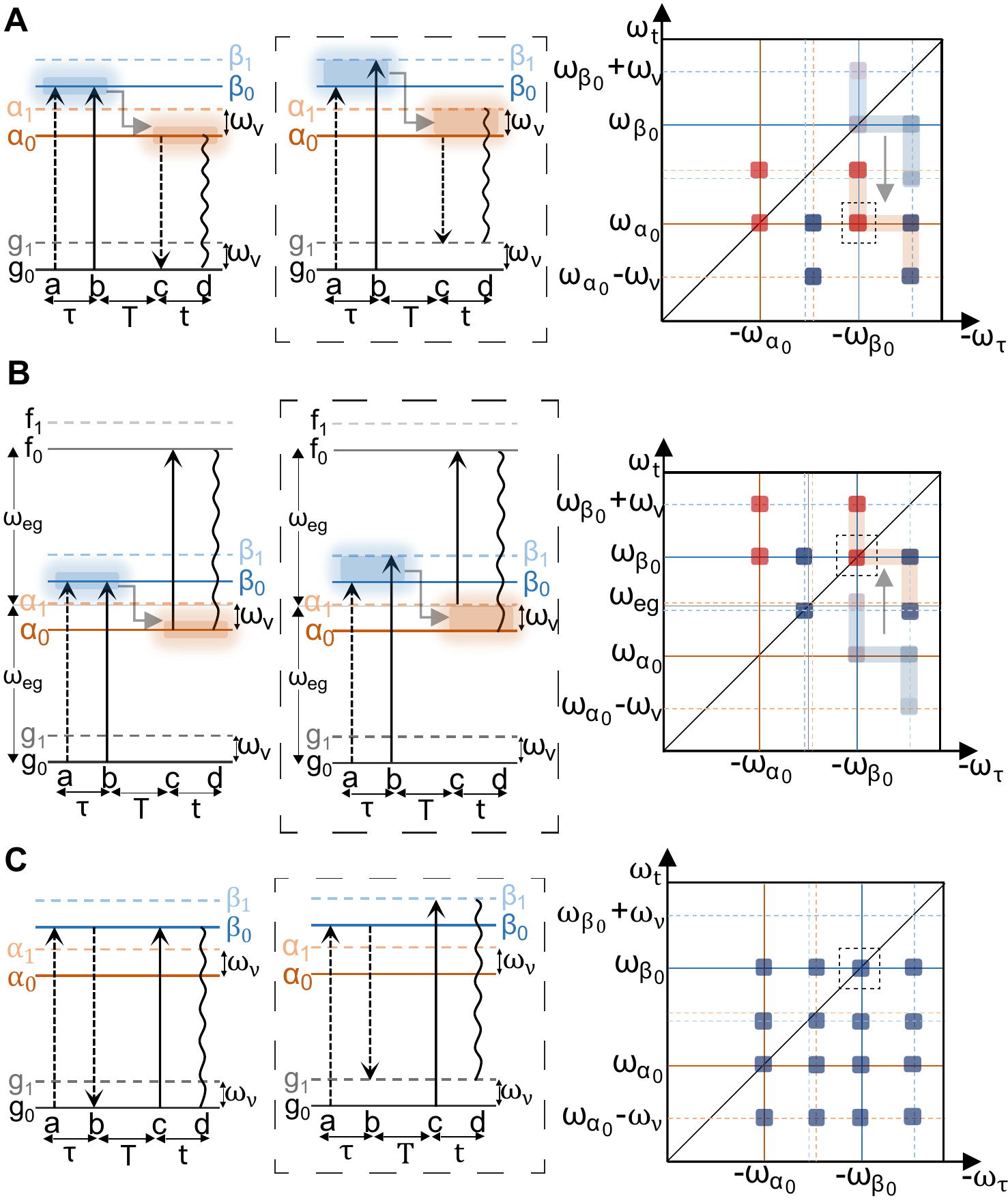}
	\caption{\footnotesize Wavemixing pathways with electronic relaxation through population and coherence transfer. (A) ESE wavemixing pathway corresponding to (left) population relaxation and (middle) vibrational coherence transfer from the donor $\beta$ to acceptor $\alpha$ excitons. (right) Expected 2D CM locations of 0-1 vibrational coherences arising from real rephasing ESE pathways. Red and blue squares denote the quantum beat phase $+\omega_T$ and $-\omega_T$, respectively. Subscripts 0,1 denote vibrational levels separated by frequency $\omega_v$.  Length of first and last arrow in the wavemixing pathways determines the 2D location along the excitation and detection axis, respectively. The arrow marks the chair shift from $DP_U$ to $CP_L$ due to vibrational coherence transfer. The wavemixing diagram for the peak highlighted by the dashed square is shown in the middle figure, all other pathways are shown in Fig.~S6. (B) ESA wavemixing pathways and CM contributions. The CM chair shifts opposite to ESE, from $CP_L$ to $DP_U$ (C) GSB wavemixing pathways and CM contributions. From the 2D CMs, it is evident that after vibrational coherence transfer, CM contributions on $DP_U$ will be a result of interfering GSB and ESA pathways.}
	\label{fig:fig2}
\end{figure*}
\FloatBarrier 

Fig.~\ref{fig:fig2} connects wavemixing diagrams incorporating ultrafast $D \rightarrow A$ population relaxation and vibrational coherence transfer to corresponding contributions on the 2D spectrum. It is well understood\cite{Engel2022} that the 2D locations of ESE and ESA population contributions on \add{$DP_U$} and lower cross-peak ($CP_L$), respectively, will be interchanged by population transfer. This is illustrated in the wavemixing pathways in the first column in Figs.~\ref{fig:fig2}A-C. However, it is vital to recognize the corresponding evolution of coherence pathways after vibrational coherence transfer. This is shown in the wavemixing pathways in the middle column, with corresponding 2D CM contributions in the right column. Interestingly, due to vibrational coherence transfer dominant along spectator modes such as $\hat{q}_+$ in an excitonic dimer, a concomitant shift in the positions of coherence peaks is also expected. 0-1 vibrational coherences contribute as a chair pattern\cite{Seibt2013} in a 2D CM, such that the entire chair pattern of ESE/ESA coherence peaks is interchanged as well. Only wavemixing pathways for one of the coherence peaks in shown and marked as dashed square in the 2D CM. \AS{All other contributions are shown in Fig.~S6}. Recalling monomer CM lineshapes in Eqns.~\ref{eq3}--\ref{eq4}, vibrational coherence transfer in an excitonic dimer implies ESE vibrational coherence pathways on $DP_U$ are replaced by ESA, which interfere with unshifted GSB pathways to result in $\beta(-\omega_{\tau},\omega_t)$ nodal lineshapes, with the extent of destructive interference on the 2D diagonal dependent on ESA strength $\kappa$. 

In order to confirm the above expectations, we simulate 2DES spectra which include exact non-adiabatic couplings through numerically diagonalized eigenvectors, phenomenological population relaxation and coherence transfer through phenomenological relaxation incorporated in sum-over-states response functions\cite{Jonas2003}, optical decoherence through Brownian oscillators and ensemble dephasing\cite{JonasARPC2018} of vibronic coherences through energetic averaging. The vibrational frequencies and weak FC displacements are based on intramolecular FC active vibrations of \textit{BChl a}. In the study of Palec\v{e}k et al.\cite{Palecek2017}, the $B-H$ exciton energy gap of 650 cm$^{-1}$ was in vibronic resonance\cite{Tiwari2013} with a prominent intramolecular vibrational frequency of \textit{BChl a}. Enhanced\cite{Tiwari2013} GSB vibrational coherences arising from this resonance were previously reported\cite{Ryu2014} by Ryu et al. To study the effect of vibronic resonance on the expected destructive interference, the diabatic exciton energy gap is chosen to be resonant with a 650 cm$^{-1}$ intramolecular vibration, while the other vibrational frequency of 350 cm$^{-1}$ does not participate in resonant vibronic mixing. Choosing a resonant and non-resonant vibrational mode provides a minimum model to explain the diagonal nodal lines reported\cite{Palecek2017,Policht2022} for \textit{all} observed intramolecular vibrations. Note that we choose complete two-particle basis sets in our calculations such that vibronic enhancement and multiple wavemixing pathways\cite{TiwariThesis} due to resonance are accurately captured in the simulations. Describing multiple wavemixing pathways arising at resonance is necessary to assess whether they appreciably perturb the destructive interference on the 2D diagonal. The anti-correlated energetic disorder of 68 cm$^{-1}$ is chosen such that ensemble dephasing\cite{JonasARPC2018} of purely electronic coherences is complete within $\sim$200 fs and consistent with experiments\cite{Palecek2017}.

Following previous approaches\cite{Jonas2008b,Policht2022} treating vibrational coherence transfer phenomenologically through Feynman pathways, the model here does not consider the details of ultrafast energy transfer mechanism. Instead population transfer timescale of $\sim$ 50 fs \cite{Jonas1996,Niedringhaus2018} is included phenomenologically, such that the energy transfer is near complete\cite{Palecek2017} by $\sim$200 fs. $\mathcal{\textbf{R}}_{\alpha \beta = \gamma \delta}$ coherence transfer is also included phenomenologically in the analytic response functions (Section S3). \vt{The simulations are carried out in the site basis with intramolecular vibrational coordinates $\hat{q}_{A,B}$ which allow for both correlated $\hat{q}_+$ and anti-correlated $\hat{q}_-$ vibrational motions on the donor and acceptor excitons. To understand whether vibronic resonance affects the diagonal nodal line, calculations along only the spectator mode $\hat{q}_+$ where no resonant vibronic mixing is possible\cite{Tiwari2017}, are also analyzed. Vibronic coherences are not expected to survive anharmonic non-adiabatic couplings\cite{Peters2017}. Experimentally, dephasing rates for such wavepackets \vt{are not known} in case of RCs.  For a photosynthetic antenna, Thyrhaug et al. have reported\cite{Thyrhaug2018} $\sim$4-5x faster dephasing of vibronic versus purely vibrational wavepackets. In our simulations, we do not include any excited state dephasing for vibronic wavepackets. This allows us to infer the effect of surviving vibronic coherences, if any, on the expected nodal line.} \\

Because symmetric vibrations in an excitonic dimer maintain a fixed donor-acceptor energy gap, they do not participate\cite{Tiwari2013} in vibronic mixing, such that, in the simulations, vibronic eigenvectors with $\hat{q}_+$ excitations undergo vibrational coherence transfer. In contrast, vibronic eigenvectors with only $\hat{q}_-$ excitations strongly mix excitons near a vibronic resonance to result in energetic splittings (for example, see Fig.~1 of ref.~\cite{Tiwari2018}). Consequently, the corresponding vibronic wavepackets in the simulations do not undergo coherence transfer. A multiplicative factor $\kappa$ is included in the ESA response as a parameter to account for the collective ESA transition strength. Fig.~S2 and Fig.~S6 of ref. \cite{Niedringhaus2018} suggest that ESA contributions on the main diagonal peak are already significant by $T = 250$ fs. Similar to experiments, the CMs are calculated from $T =$ 200 fs after dephasing of electronic coherence and energy transfer are approximately complete. All the model parameters are described in \AS{Section S3}.

\begin{figure*}[h!]
	\centering
	\includegraphics[width=3.2 in]{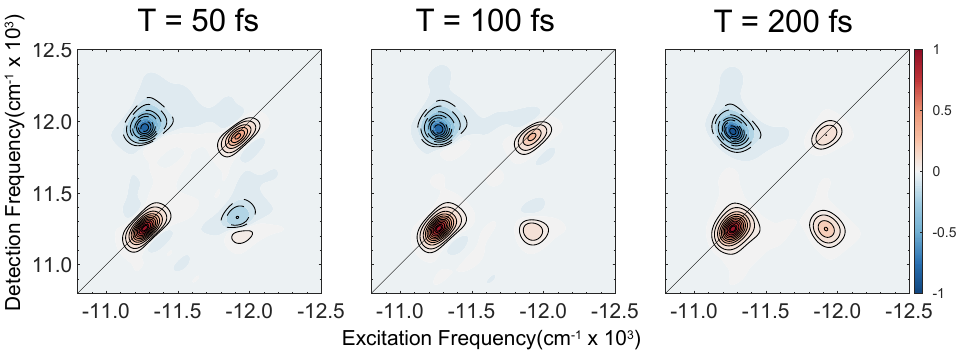}
	\caption{\footnotesize Real absorptive 2D spectra for waiting times $T =$ 50, 100 and 200 fs with an ESA signal strength of $\kappa = 2$ and temperature of 77 K. The simulations details and model parameters are described in Section S3.}.
	\label{fig:fig3}
\end{figure*}
\FloatBarrier

\begin{figure*}[h!]
	\centering
	\includegraphics[width=3 in]{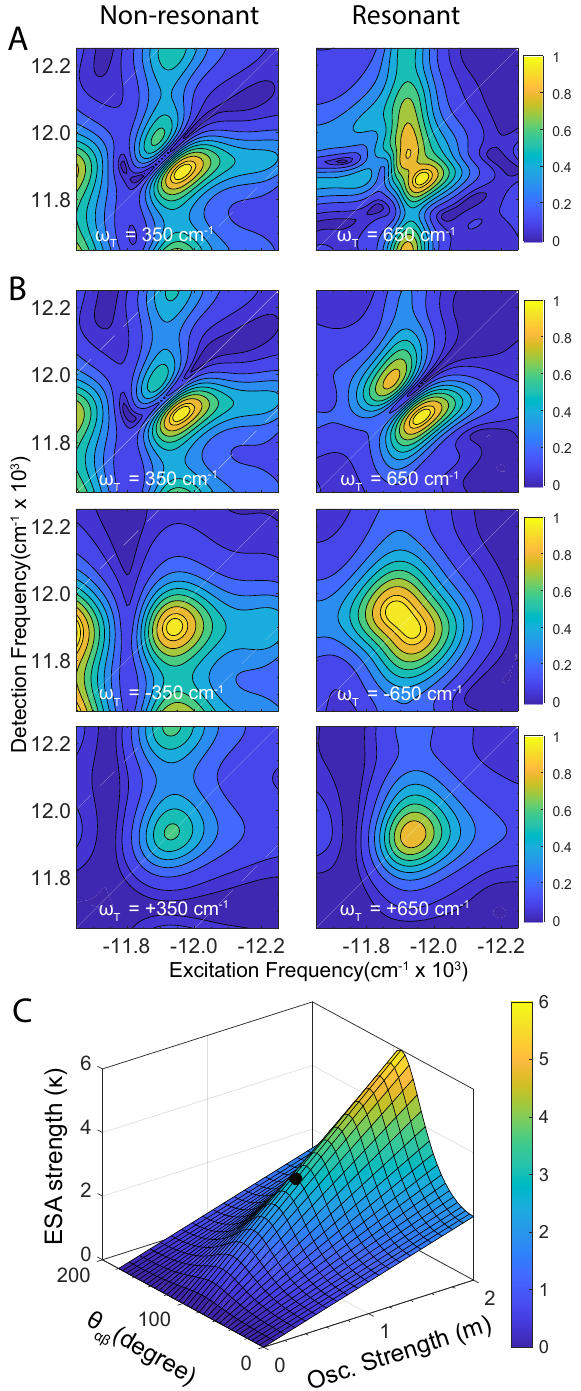}
	\caption{\footnotesize (A)Zoomed $DP_U$ real rephasing 2D CM for non-resonant (350 cm$^{-1}$) and resonant (650 cm$^{-1}$) vibration. The calculation corresponds to the case where both $\hat{q}_-$ and $\hat{q}_+$ tuning and spectator modes, respectively, are allowed.  (B) 2D CM calculation with only spectator mode $\hat{q}_+$ (top), and resolved into -$\omega_T$  (middle) and +$\omega_T$ (bottom) components. The full CMs with the red square around $DP_U$ are shown in \AS{Figs.~S7--S8}.  Contours are drawn at the 5$\%$, and 10-90$\%$ in 10$\%$ intervals. (C) The surface plot of ESA signal strength $\kappa$ required for complete destructive interference between GSB-ESA coherence pathways, as a function of the relative excitonic oscillator strength ($m$) and the angle between the excitonic transition dipoles ($\theta_{\alpha\beta}$). The detailed strength calculation is shown in \AS{Section S4}. Black dot shows required $\kappa$ for perpendicular and equal magnitude transition dipoles.}
	\label{fig:fig4}
\end{figure*}

Fig.~\ref{fig:fig3} shows the total absorptive 2D spectra at $T =$ 50, 100 and 200 fs using the above model. As expected from the analysis in Fig.~\ref{fig:fig2}, population relaxation between excitons causes $DP_U$ and $CP_L$ ESE population signals to decay and grow, respectively, with an opposite trend for ESA signals. The total 2D signal at $DP_U$ is diminished due to the cancellation of positive GSB and negative ESA signals with increasing waiting time $T$, consistent with experiments which report\cite{Niedringhaus2018} a dominant ESA signal on $DP_U$. The corresponding 2D CMs are shown in Fig.~\ref{fig:fig4} for the resonant 650 cm$^{-1}$ and the non-resonant 350 cm$^{-1}$ vibration. The CMs are zoomed in on $DP_U$ to highlight the expected destructive interference on the 2D diagonal due to vibrational coherence transfer along the spectator modes. Zoomed out CMs are shown in \AS{Figs.~S7--S8}. Fig.~\ref{fig:fig4}A compares the $DP_U$ diagonal node for a resonant versus non-resonant vibration. The calculation assumes that excited state vibronic wavepackets along the tuning mode $\hat{q}_-$ survive energy transfer. Both vibrations show a nodal line on $DP_U$ due to GSB-ESA destructive interference. However, the destructive interference for the resonant vibration is strongly perturbed by unshifted ESE vibronic coherence contributions due to lack of coherence transfer along $\hat{q}_-$. This is accentuated further by multiple wavemixing pathways contributing on $DP_U$ at resonance (\AS{Section S5}). In contrast, Fig.~\ref{fig:fig4}B shows a calculation only along the spectator mode $\hat{q}_+$ which does not participate\cite{Tiwari2013} in vibronic mixing.  As expected, a clean nodal line consistent with experimental reports is seen on $DP_U$ due to destructive interference of GSB-ESA coherent pathways following vibrational coherence transfer. These findings suggest that vibronic wavepackets either do not survive energy transfer or there is only weak resonant vibronic mixing in case of \add{BRCs} due to sub-peaks within the $B$ and $H$ bands in \add{BRCs}\cite{Rancova2016} and narrow vibronic resonance widths\cite{Tiwari2017,Tiwari2018}. The former seems plausible given the recent observations\cite{Thyrhaug2018} of Thyrhaug et al. of $\sim$4-5x faster dephasing of vibronic versus vibrational wavepackets in a photosynthetic antenna. The last two rows in Fig.~\ref{fig:fig4}B show that, as expected, when CMs are resolved according to quantum beat phase along $\omega_T$, destructive interference on the 2D diagonal does not occur. Note that the impulsive 2D calculation presented here allows all CM peaks to be prominent. Laser pulse frequency filter\cite{Jonas2003} from a finite pulse is expected to suppress off-diagonal CM peaks.
The above model and calculations have assumed perpendicular and equal magnitude excitonic transition dipoles to preclude any transition dipole interference effects. However, transition dipoles in RCs are not perpendicular or of equal magnitude.  Below we consider whether interference between non-orthogonal unequal magnitude transition dipoles in RCs can cause nodal line feature to appear readily. We have derived an analytic expression for ESA strength $\kappa$ required for complete destructive interference on $DP_U$, as a function of excitonic transition dipole angle $\theta_{\alpha\beta}$ and their relative oscillator strength ratio $m$. The derivation is described in Section \AS{S4}, with final expression given by,
\begin{eqnarray}
	\kappa =\frac{ 3m}{1+2 \cos[2](\theta_{\alpha\beta})},
	\label{eq5}
\end{eqnarray}
and plotted in Fig.~\ref{fig:fig4}C. An ESA strength of $\kappa =$ 3 is required for complete destructive interference in case of perpendicular and equal magnitude transition dipoles. \AS{Table S4} calculates the transition dipole strengths and directions expected\cite{Fleming2001, Jonas1996} from $P$, $B$ and $H$ excitons in the \add{BRCs}. Using Eqn.~\ref{eq5} and Table S4, in case of BRCs the ESA strength required for GSB-ESA destructive interference is reduced to $\kappa\sim2$ for both $P-B$ and $B-H$ exciton pairs. This seems reasonable as a collective ESA strength given the multiple two-quantum electronic manifolds in the RCs and the experimental observations (Fig.~S2 and Fig.~S6 of ref.\cite{Niedringhaus2018}). \AS{Fig.~S10} compares the nodal feature in Fig.~\ref{fig:fig4}B for ESA strengths of $\kappa = $1--3. \add{The insensitivity of destructive interference to $\kappa$ suggests that the nodal feature, although imbalanced, is expected to be persistent over a range of ESA strengths. Several additional effects such as peak overlaps, inhomogeneous broadening, laser pulse filter, and red- or blue-shifted ESA signals \add{due to electrochromic shifts} can also cause an imbalanced nodal feature.  For example, Fig.~S11 shows that $DP_U$ node becomes imbalanced, although persistent, even when expected\cite{Zigmantas2021} BRC electrochromic shift of $\sim$100-200 cm$^{-1}$ is included.} \\

\FloatBarrier 

\add{Coherence transfer along spectator modes is expected to accompany ultrafast electronic energy transfer. Our simulations of excitonic dimer model and Feynman pathways account for this phenomenologically without considering the details of energy transfer or multiple 1- and 2-quantum electronic manifolds in the RCs. This phenomenological treatment of coherence transfer between vibronic eigenstates is consistent with that recently proposed by Policht et al.\cite{Policht2022} to explain coherence contributions in the upper cross-peak region arising from ESA pathways. In our case, by accounting for shifting population and coherence contributions in the 2D spectra, we have shown that the resulting GSB--ESA interference between coherent pathways contributing on the 2D diagonal produces persistent signatures, similar to those reported in BRC 2DES studies\cite{Niedringhaus2018} for all observed intramolecular vibrations.} 


\section*{Conclusions}

\add{Our results provide new insights on the connections between vibrational coherence transfer, interference between coherent Feynman pathways and the resulting 2D CM lineshapes. By establishing these connections we resolve the distinct physical origins that lead to reportedly\cite{Palecek2017,Policht2018,PolichtThesis} similar nodal diagonal lineshapes in the 2D CMs of two disparate systems, \textit{BChl a} monomer and multichromophoric \add{BRCs}. We show that while the former arises from unique phase-twists from interfering GSB-ESE coherence pathways, the latter is fundamentally different and arises from previously overlooked vibrational coherence transfer along spectator modes accompanying ultrafast electronic relaxation. By incorporating relaxation pathways in Feynman diagrams, we show that along with population transfer, vibrational coherence transfer leads to a concomitant shift of coherence contributions on the 2D CM spectrum. Such coherence-shifts are expected to be dominant along spectator modes. The resulting destructive interference between GSB--ESA signal pathways is consistent with the reported nodal lines on the 2D diagonal. From experimental observations of dominant ESA signals, and estimations of dipole strengths and directions in \add{BRCs}, our analysis suggests that such nodal lines may be readily expected in \add{BRCs}. Our results resolve recent spectroscopic observations, highlight the rich information content of a 2D CM spectrum, and establish its usefulness as a subtle spectroscopic reporter of underlying electronic relaxation mechanisms.}

\section*{Acknowledgments}
AS acknowledges research fellowship from the Indian Institute of Science (IISc). This project is supported by Science and Engineering Research Board, India under grant sanction number CRG/2019/003691 and Department of Biotechnology, India under grant sanction number BT/PR38464/BRB/10/1893/2020. 
 

\section*{Supporting Information Available}

Bloch model analytical calculations, monomer 2D simulation parameters, 2D simulations with population and coherence transfer, simulations for ESA strength dependence of nodal line, multiple sub-peaks on upper 2D diagonal at vibronic resonance, analytical expression for nodal line with generalized transition dipole directions and magnitudes, 2D and CM calculations with blue shifted ESA signal, Figures S1--S11.

\bibliography{VibNodeRefs}
\bibliographystyle{achemso}

\end{document}